\documentclass[aps,prl,twocolumn,epsfig,amsmath,floatfix,]{revtex4}

\usepackage{xcolor}
\usepackage{epsfig,amsmath}
\usepackage{graphicx}% Include figure files
\usepackage{dcolumn}% Align table columns on decimal point
\usepackage{bm}% bold math %\bibliographystyle{apsrev}
\usepackage[]{amsmath}
\usepackage{braket}
\usepackage{amsfonts,amsmath}
\usepackage{epsfig,amsmath}
\usepackage{graphicx}% Include figure files
\usepackage{dcolumn}% Align table columns on decimal point
\usepackage{bm}% bold math %\bibliographystyle{apsrev}
\usepackage{bbold}
\newcommand{\beq}{\begin{equation}}
\newcommand{\eeq}{\end{equation}}
\newcommand{\beqa}{\begin{eqnarray}}
\newcommand{\eeqa}{\end{eqnarray}}

\newcommand{\la}{\langle} 
\newcommand{\ra}{\rangle}

\def\nphy#1{{Nature\ Phys.} {\bf#1}}
\def\natphot#1{{ Nat.\ Phot.} {\bf#1}}
\def\njp#1{{ New\ J.\ Phys.} {\bf#1}}

\def\ol#1{{ Opt.\ Lett.} {\bf#1}}

\def\pra#1{{ Phys.\ Rev. A\/} {\bf#1}}

\def\prl#1{{ Phys.\ Rev.\ Lett.} {\bf#1}}

\begin{document}

\title{Super-resolution of two unbalanced point sources assisted by the entangled partner}
\author{Abdelali Sajia}
\author{Xiao-Feng Qian}
\email{xqian6@stevens.edu}
\affiliation{Center for Quantum Science and Engineering, and Department of Physics, Stevens Institute of Technology, Hoboken, New Jersey 07030, USA}



\begin{abstract}
Sub-diffraction-limit resolution, or super-resolution, had been successfully demonstrated by recent theoretical and experimental studies for two-point sources with ideal equal-brightness and strict incoherenceness. Unfortunately, practical situations of either non-equal brightness (i.e., unbalancenss) or partial coherence are shown to have fatal effects on resolution precision. As a step toward resolving such issues, we consider both effects together by including an entangled partner of the two-point sources. Unexpectedly, it is found that the two negative effects can counter affect each other, thus permitting credible super-resolution, when the measurement is analyzed in the entangled partner's rotated basis. A least resolvable finite two-source separation is also identified analytically. Our result represents useful guidance towards the realization of super-resolution for practical point sources. The vector-structure analog of quantum and classical light sources also suggests that our analysis applies to both contexts. 
\end{abstract}

\maketitle

%%%%%%%%%%%%%%%%%%%%%%%%%%  body  %%%%%%%%%%%%%%%%%%%%%%%%%%
\section{Introduction}

Resolution of two point sources is one of the most crucial elements in the science of imaging and sensing. The quality of fine resolution relies on two major factors: resolution capability (how small a separation can be discriminated) and measurement estimation credibility (how much the measurement can be trusted). For over a century, the empirical Abbe-Rayleigh diffraction criterion \cite{Abbe1873, Rayleigh1896, BornWolf,Rayleigh1979}, relating to the ratio of light wavelength and aperture diameter, has been regarded as a roadblock that limits the resolution capability with practically feasible parameters \cite{Goodman2005,Kolobov2000PRL}. This is due to the fact that when the two sources are getting closer, the blurred overlapping signals are harder to discriminate through direct intensity measurements. Moreover, the second factor, measurement credibility (or measurement precision), will also reduce as noise effects become relatively more prominent when the separation of the two sources decreases. Studies have shown that while statistical methods were able to improve the resolution capability by determining the source locations, the precision of measurement vanishes as the separation of the two sources goes beyond the diffraction limit, approaching zero \cite{Bettens1999UM, Aert2002JSB, Ram2006PNAS, Chao2016JOSA}. This phenomenon consolidates the common wisdom of the Abbe-Rayleigh diffraction criterion from a more rigorous foundation, and is termed as the Rayleigh's curse by many authors, see for example in \cite{Tsang2015O, Tsang2016PRL, Tsang2016PRX, Paur2016O, Rehacek2017PRA, Larson2018O, Hradil2019O, Liang2021O, Wadood2021OE,Faber2000OL}. 

Recently, the pioneering works of Tsang and coworkers \cite{Tsang2015O, Tsang2016PRX,Tsang2019arxiv,Tsang2019O} showed that it is possible in principle to improve both factors by analyzing the signal in a different (e.g., the Hermite-Gaussian mode) spatial basis instead of direct intensity measurement. The new technique leaves the Abbe-Rayleigh diffraction criterion irrelevant and at the same time guarantees a finite desired estimation accuracy via the parameter Fisher information (FI) \cite{Helstrom1976, Holevo2001}. Experimental confirmations have also been demonstrated, see for examples in \cite{Paur2016O, Wadood2021OE}. While working perfectly in ideal situations, this technique has two constraints, requiring (1) incoherence and (2) balance (equal brightness) of the two point sources. It has been shown that releasing either one of the two may lead to the resurgence of Rayleigh’s curse by Rehacek et al., with unbalanced incoherent sources \cite{Rehacek2017PRA} and by Larson-Saleh \cite{Larson2018O, Larson2019O} and De, et al., \cite{DePRR2021} with balanced but partially coherent sources .

\begin{figure}[h!]
\centering
\includegraphics[width=4cm]{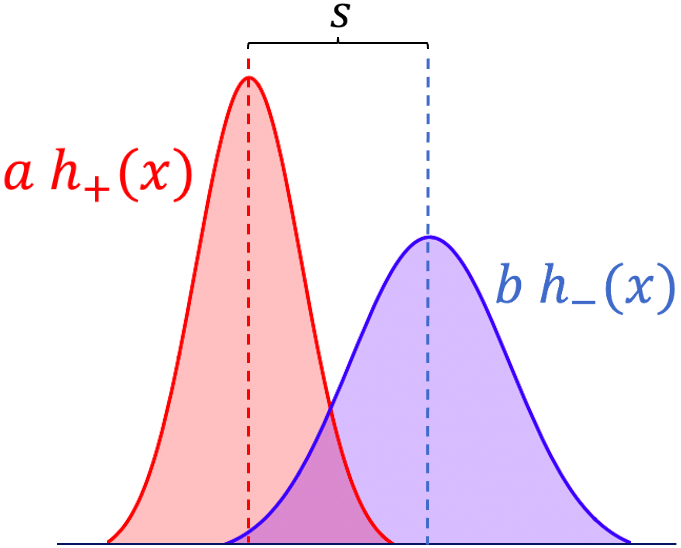}
\caption{The point spread functions, $h_{+}(x)$ and $h_{-}(x)$, of two unbalanced point sources separated by $s$ via a shift-invariant imaging system. The coefficients $a$ and $b$ characterize a continuous unbalanceness.}
\label{scheme}
\end{figure}

To address this issue, we attack both restrictions at the same time by investigating the resolution of two unbalanced and partially coherent point sources with the assistance of an entangled partner, see a schematic illustration in Fig.~\ref{scheme}. The effect of partial coherence is analyzed by continuous basis rotation of the entangled partner \cite{Plenio2017RMP}. It is found that the effect of unbalanceness on the two-source separation estimation parameter, i.e., Fisher Information, is equivalent to that of the basis rotation-induced partial coherence. Unexpectedly, the joint effect of the two restrictions by entanglement permits the realization of super-resolution with a finite Fisher Information even when the separation of the two sources is approaching zero. It is also determined analytically that there exists a ``least resolvable" (non-zero) distance of which the Fisher information experiences a non-zero minimum that is determined by Lambert W functions \cite{Corless1996ACM}. This provides a guidance to employ optimum practical parameters to achieve super-resolution with a given accuracy requirement.

\section{Model and Methods}

We consider two unbalanced sources located at $x\pm s/2$ respectively with a separation $s$. Through a shift-invariant imaging system, their normalized point spread functions (PSF) are assumed to take the Gaussian form, i.e., $h_{\pm}(x)=h(x\pm s/2)$ with $h(x)=\frac{1}{\sqrt{2\pi\sigma^{2}}}\exp[{-\frac{x^{2}}{2\sigma^2}}]$ and $\sigma$ as the width, see an illustration in Fig.~\ref{scheme}. We include the partial coherence property of the two sources by introducing an entangled partner of the spatial degree of freedom. The optical field in the image plane can be described as 
\begin{align}
\label{ent}
 \ket{\Psi_{ent}}&=a\ket{h_{+}}\ket{\phi_{1}}+be^{i\varphi}\ket{h_{-}}\ket{\phi_{2}},
\end{align}
where $\ket{h_{\pm}}$ represent the two non-orthogonal point spread functions (vectors) of the spatial degree of freedom with $\la x \ket{h_{\pm}}=h_{\pm}(x)$, $\ket{\phi_{1}}, \ket{\phi_{2}}$ are two generic states describing the remaining degrees of freedom (including polarization, temporal domain, etc.), $a$ and $b$ are real normalized coefficients with $a^{2}+b^{2}=1$, and $\varphi$ is an arbitrary phase. The degree of unbalanceness of the two sources can be simply quantified by the ratio $r=b/a$, where $r=0$ means completely unbalanced and $r=1$ indicating balanceness ($r>1$ is equivalent to $r<1$ since $a$ and $b$ are symmetric.  Here the spatial state space $\{\ket{h_{+}}, \ket{h_{-}}\}$ is entangled with the remaining state space $\{\ket{\phi_{1}}, \ket{\phi_{2}}\}$. The entangled state (\ref{ent}) can either be a quantum state of single photons or a macroscopic classical optical field \cite{Spreeuw1998FP, Luis2009OC, Simon2010PRL, Borges2010PRA, Qian2011OL, Kagalwala2013NP, Toppel2014NJP, Zela2014PRA, Forbes2017Nat, Goldberg2021AOP}.
% therefore they can be written as  $a=\cos\eta$ and $b=\sin\eta$.

In general, $\ket{\phi_{1}}, \ket{\phi_{2}}$ can be non-orthogonal and their overlap, $\la \phi_1|\phi_2\ra$, determines the degree of coherence. For the convenience of following discussions, we take the two to be orthogonal but analyze the operations (or detections) in an arbitrarily rotated basis, i.e.,  $|\phi^{\alpha}_1\ra=\cos\alpha |\phi_1\ra-\sin\alpha |\phi_2\ra$ and $|\phi^{\alpha}_2\ra=\sin\alpha |\phi_1\ra+\cos\alpha |\phi_2\ra$. This is an equivalent way of introducing partial coherence due to its basis-dependent nature \cite{Plenio2017RMP, Qian2011OL, Kagalwala2013NP, Goldberg2021AOP}, see an illustrative analysis in supplemental material section 1. 

The optical field can be rewritten in the new basis as
\begin{align}
\ket{\Psi_{ent}}=\ket{h_{1}}|\phi^{\alpha}_1\ra+\ket{h_{2}}|\phi^{\alpha}_2\ra,
\label{rotated}
\end{align}
where $\ket{h_{1}}=a\cos\alpha\ket{h_{+}}-b\sin\alpha e^{i\varphi}\ket{h_{-}}$, $\ket{h_{2}}=a\sin\alpha\ket{h_{+}}+b\cos\alpha e^{i\varphi}\ket{h_{-}}$ are two new spatial functions that are in general non-normalized and non-orthogonal.

To quantify the likely-hood of being an accurate estimation (or the degree to which measurements can be trusted) of the two-source separation $s$, the conventional Fisher information (FI) \cite{Helstrom1976, Pang2017NC, Shukur2020Nat} is employed. It's definition is based on the Cram\'er-Rao bound \cite{Zmuidzinas2003OSA, Kay1993, Pang2014PRL} ${\rm Var}(s)\geq 1/F$. The optimal estimation of the unknown parameter $s$ corresponds to the maximization of the Fisher information $F$ which corresponds to a minimum of the estimator variance ${\rm Var}(s)$.

A recent debate about the nonphysical divergence of Fisher information \cite{Larson2018O, Tsang2019O, Larson2019O, Hradil2019O} suggests that the Fisher information for analyzing two-source super-resolution needs to be appropriately normalized. Here we adopt the approach proposed by Hradil et al. \cite{Hradil2019O}, to account the total FI as a sum of weighted components for all probabilistic events. For the general entangled state (\ref{rotated}), the FI is defined as 
\beq
F_{tot}=\la h_1|h_1\ra F_{\rho_1}+\la h_2|h_2\ra F_{\rho_2},
\label{FIdefinition}
\eeq
where $\rho_1=|h_1\ra \la h_1| /N_1$, $\rho_2=|h_2\ra\la h_2|/N_2 $ are two corresponding normalized states, by factors $N_1$, $N_2$, of the spatial domain with corresponding weights $\la h_1|h_1\ra$, $\la h_2|h_2\ra$. Here Fisher information takes the form $F_{\rho}=2{\rm Tr}[(\partial_s\rho(s))^2]$ for an arbitrary normalized state $\rho (s)$. This measure is based on the conditional outcome of a measurement in the rotated basis $\{|\phi^{\alpha}_1\ra,|\phi^{\alpha}_2\ra\}$ of the entangled partner. 

It is important to note that here we treat the measurement of signal as a single repetition (e.g., an independent single photon detection event, detecting a bunch of identical photons within the coherence time, or a single measurement of light intensity). For given number (e.g., $N$) of multiple repetitions of measurement, our results will remain the same up to the constant factor $N$. Our analysis doesn't cover environment-induced loss cases where the photon numbers are unknown.

\section{Results and Messages}

\begin{figure}[t]
\includegraphics[width=6cm]{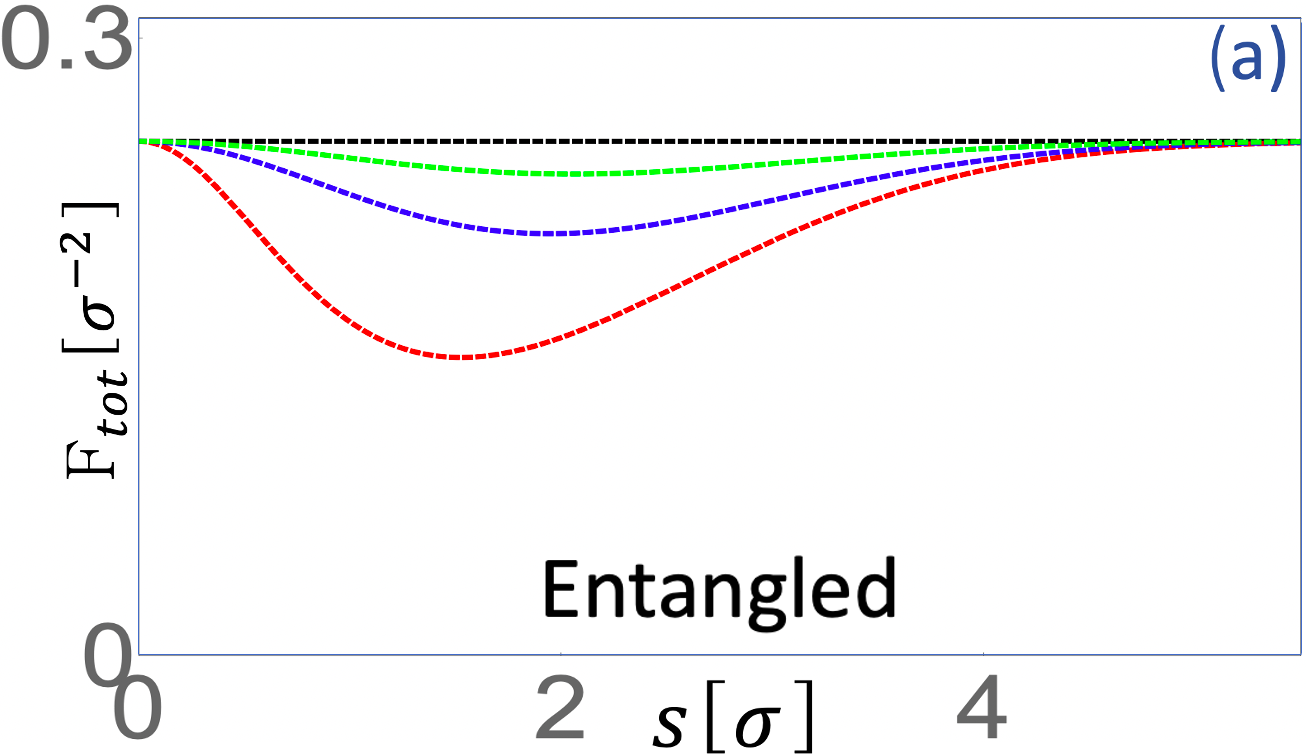}
\includegraphics[width=6cm]{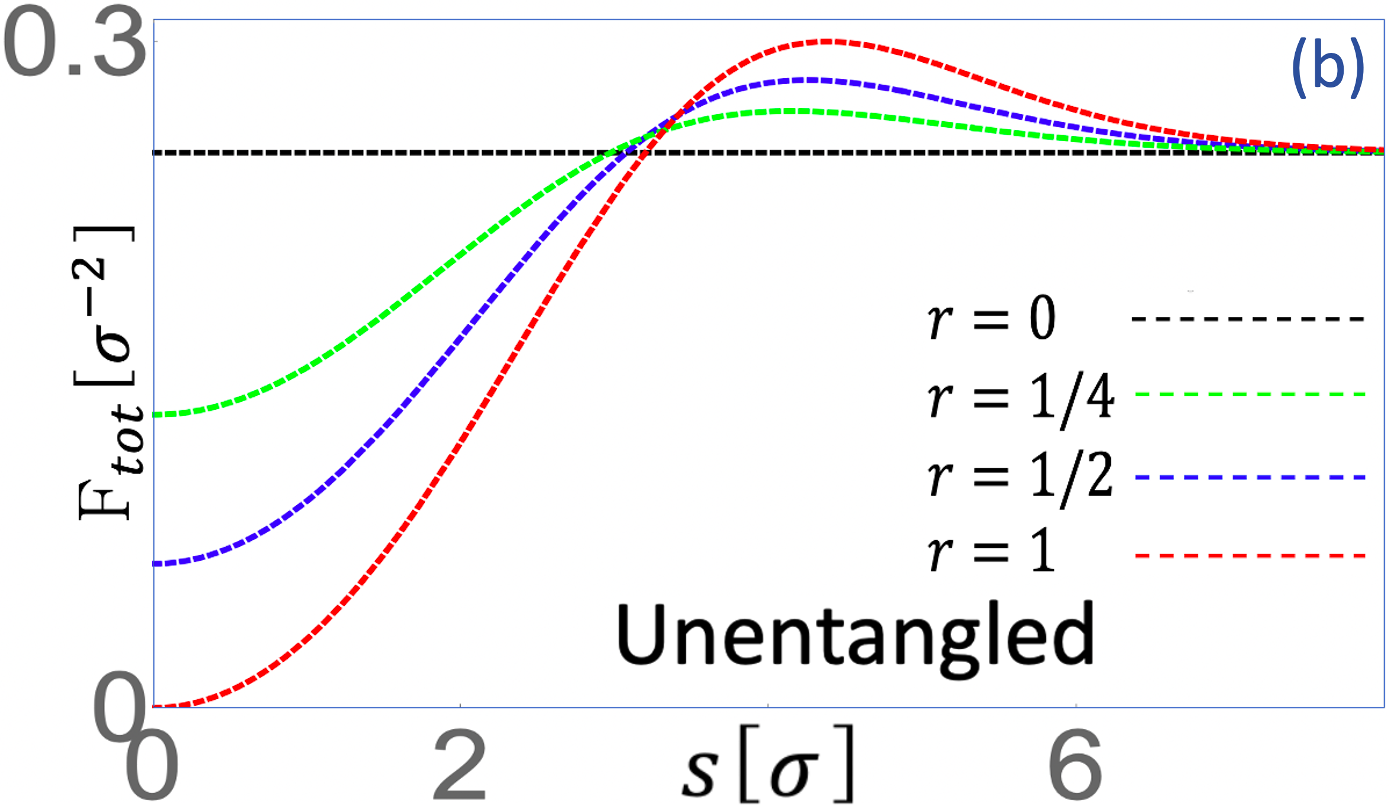}
\caption{Dependence of the FI on the displacement s for $\varphi=0$ with $\sigma=1$. Different color represents different value of $r$. The red line is for the balanced case ($r=1$), and the blue and green lines corresponds  to $r=\frac{1}{2}$, and $r=\frac{1}{4}$ respectively. For $r=0$ are represented by the black line. (a) FI for the state $\ket{\Psi_{1}}$ in the case $\alpha=\frac{\pi}{6}$. (b) FI is for the state $\ket{\Psi_{2}}$.}
\label{FI}
\end{figure}

In our consideration, the unbalanceness of the two sources $r=b/a$ is also an unknown parameter. However, it will be shown later that measurement in the rotated basis $\{|\phi^{\alpha}_1\ra,|\phi^{\alpha}_2\ra\}$ is always able to cancel the unbalanceness effect. Therefore, one needs only to consider the single unknown parameter $s$ for the calculation of the Fisher Information. Then the Fisher information for the entangled field (\ref{rotated}) can be explicitly obtained as 
%\begin{widetext}
\beqa
\label{Ftot}
F_{tot}(s)&=&\frac{1}{4\sigma^{2}}\notag \\
&-&\frac{(r\sin2\alpha\cos\varphi)^{2}s^{2}}{16[(\cos^{2}\alpha+r^{2}\sin^{2}\alpha)e^{s^{2}/8\sigma^{2}}-r\sin2\alpha\cos\varphi]} \notag \\
&\times& \frac{1}{(r^{2}\cos^{2}\alpha+\sin^{2}\alpha)e^{s^{2}/8\sigma^{2}}+r\sin2\alpha\cos\varphi},
\eeqa
%\end{widetext}
which depends on the two-source separation $s$, the entangled partner's measurement basis characterized by rotation angle $\alpha$, the two-source unbalanceness through the amplitude ratio $r$, and the relative phase $\varphi$. Fig.~\ref{FI} (a) illustrates the specific behaviors of the Fisher Information on $s$ for a fixed partial coherence (rotation angle $\alpha=\pi/6$) and relative phase $\varphi=0$, but for different unbalanceness $r=0,1/4,1/2,1$. Qualitative behaviors for other degrees of partial coherence (i.e., other rotation angles) are similar to Fig.~\ref{FI} (a) except for some special cases which will be discussed later. The detailed derivation of (\ref{Ftot}) is provided in supplemental material section 2.

There are three important messages delivered by the obtained Fisher information (\ref{Ftot}) for the entangled source. 

{\em The first major message} of result (\ref{Ftot}) is that super-resolution is achievable for various practical unbalanceness settings as the value of Fisher information remains finite even when the separation $s$ goes to zero, as shown in Fig.~\ref{FI} (a). This unexpected behavior is a result of the existence of the entangled partner. To have a clearer picture of effects of the entangled partner, we also analyze the Fisher information of two unentangled point sources with the same unbalanceness characterized by $r=b/a$ and relative phase $\varphi$, i.e.,
\begin{align}
\label{unent}
 \ket{\Psi'}&=(a\ket{h_{+}}+be^{i\varphi}\ket{h_{-}})|\phi\ra,
\end{align}
where $|\phi\ra$ is a generic state of the remaining degrees of freedom. In this non-entangled case, the measurement basis of $|\phi\ra$ is irrelevant to the spatial domain. Therefore, the Fisher information can be directly computed as $F_{\rho'}=2{\rm Tr}[(\partial_s\rho')^2]$ where $\rho'=|\Psi'\ra \la \Psi'|/N'$ with $N'$ being the normalization factor. It can be obtained straightforwardly as 
\begin{align}
\label{F-unent}
F_{\rho'}(s)&=&\frac{\frac{1}{4\sigma^{2}}- ab\cos\varphi e^{-\frac{s^{2}}{8\sigma^{2}}}\frac{(-s^{2}+4\sigma^{2})}{8\sigma^{4}})}{1+2ab\cos\varphi e^{-s^{2}/8\sigma^{2}}}\nonumber\\
&&-\frac{1}{4}\frac{(ab\cos\varphi e^{-\frac{s^{2}}{8\sigma^{2}}}s)^{2}}{(1+2ab\cos\varphi e^{-s^{2}/8\sigma^{2}})^{2}}.
\end{align}
Fig.~\ref{FI} (b) illustrates the specific behaviors of $F_{\rho'}$ for the nonentangled field with the same unbalanceness values and fixed phase $\varphi$. The detailed derivation of the above result is also provided in supplemental material section 3. 

By comparing the Fisher information of the two cases as illustrated in Fig.~\ref{FI} (a) and (b), one notes apparently that the entangled field has a significant enhancement for all unbalanceness values in the small separation $s$ regime. Particularly, for the frequently studied balanced source case ($r=1$), the nonentangled field Fisher information vanishes at zero separation while the entangled field one achieves its maximum finite value. It is also interesting to note that for the entangled field case, the Fisher information of various different unbalanceness simply converge to the same finite value when the source separation decreases to zero.

{\em The second major message} of result (\ref{Ftot}) lies in the underlying competing mechanism between coherence and unbalanceness in affecting the Fisher information. As is mentioned, neither partial coherence nor unbalancess is able to achieve super-resolution \cite{Rehacek2017PRA, Larson2018O}. Surprisingly, as is shown here above, the combination of the two works! This is due to the fact that coherence and unbalanceness have counter effects against each other on the Fisher information.

\begin{figure}[t!]
\includegraphics[width=6cm]{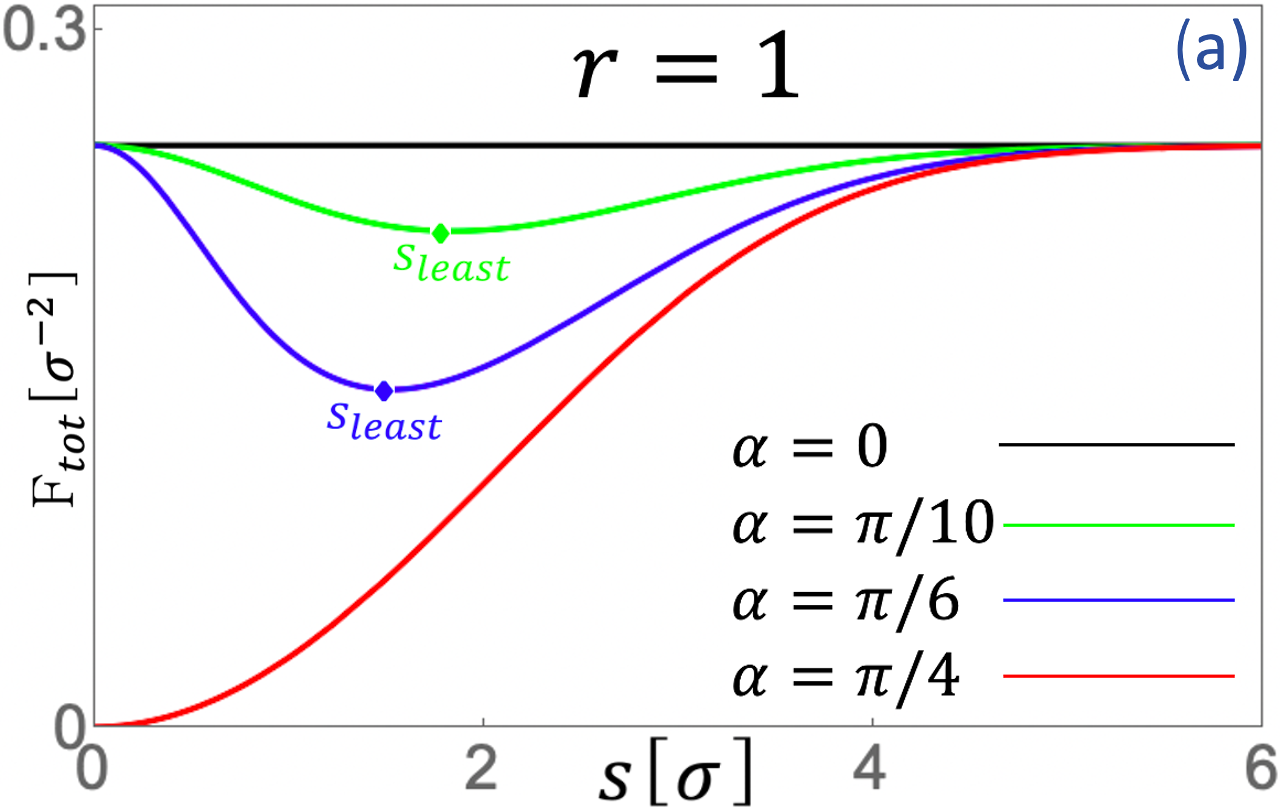}
\includegraphics[width=6cm]{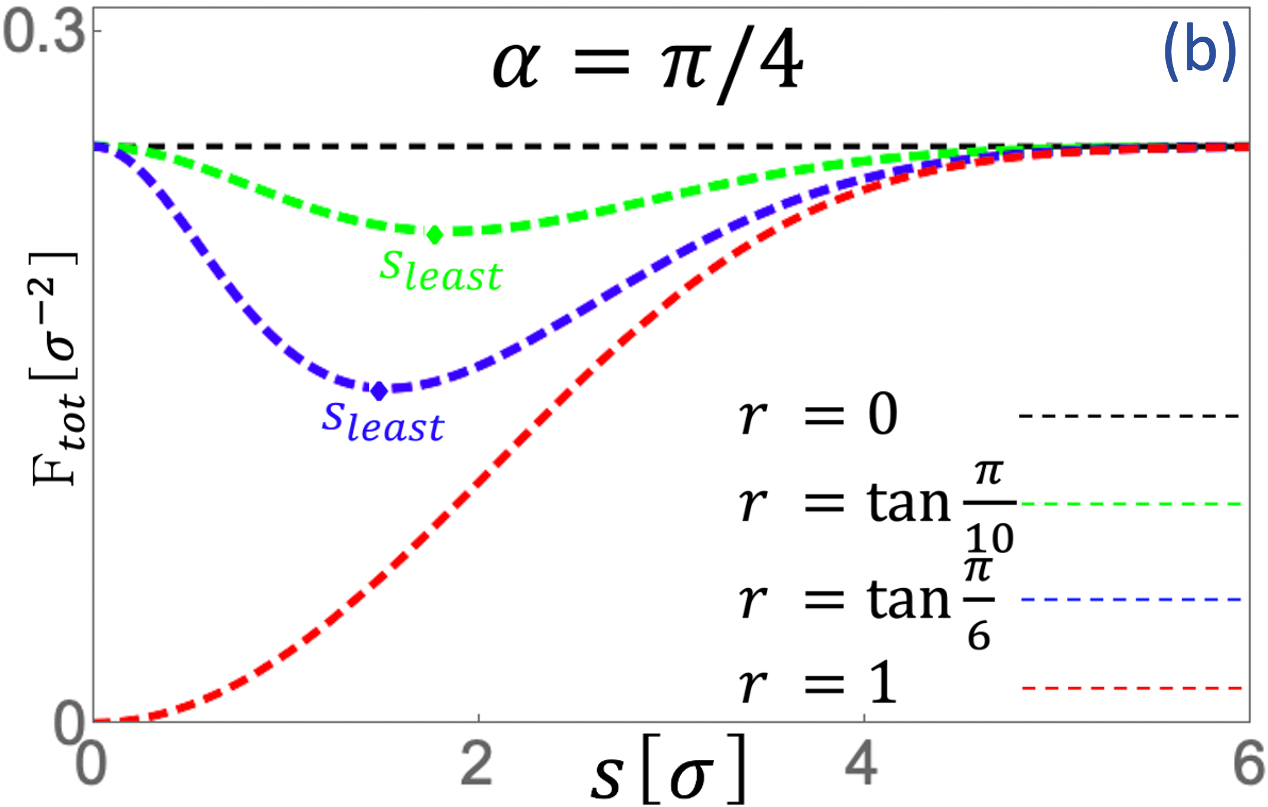}
\caption{Total FI versus the displacement $s$ for $\varphi=0$ when $\sigma=1$. (a) Rotating basis angle $\alpha$ effect on FI in the balanced case. The minimum of the FI corresponds to $s_{\rm least}$ in the expression (\ref{minimum})  (b) Unbalancenees intensities effect  in the case of $\alpha=\frac{\pi}{4}$.}
\label{equivalent}
\end{figure}

To understand this point better, we perform a detailed analysis of effects from both properties to explore the analogous behavior of the two. We first analyze the coherence effect by fixing at the two-source balanced case, i.e., setting the unbalanceness parameter to be $r=1$. Then the Fisher information simply reduces to 
\begin{align}
\label{coherence-effect}
F_{tot}(s,r=1)&=\frac{1}{4\sigma^{2}}-\frac{1}{16}\frac{\sin^{2}2\alpha\cos^{2}\varphi s^{2}}{e^{s^{2}/4\sigma^{2}}-\sin^{2}2\alpha\cos^{2}\varphi},
\end{align}
which depends on the coherence rotation angle $\alpha$ and two-source separation $s$. Fig~\ref{equivalent} (a) illustrates its behavior for different coherence parameter values of $\alpha$.

Next we analyze the unbalanceness effect by fixing the coherence rotation angle $\alpha=\pi/4$. To be comparable we define the unbalanceness parameter as $r=b/a=\tan\eta$, where $a=\cos \eta$ and $b=\sin\eta$ to satisfy the normalization condition $a^2+b^2=1$. Then the Fisher information reduces to 
\begin{align}
\label{unbalanceness-effect}
F_{tot}(s,\alpha=\frac{\pi}{4})&=\frac{1}{4\sigma^{2}}-\frac{1}{16}\frac{\sin^{2}2\eta\cos^{2}\varphi s^{2}}{e^{s^{2}/4\sigma^{2}}-\sin^{2}2\eta\cos^{2}\varphi},
\end{align}
which depends on the unbalanceness angle $\eta$ and two-source separation $s$. Fig~\ref{equivalent} (b) illustrates the behavior for different unbalanceness values of $r$ or equivalently $\eta$.

By comparing the two expressions (\ref{coherence-effect}) and (\ref{unbalanceness-effect}), one immediately notes that they are equivalent except for the change of parameter $\alpha$ into $\eta$. This shows that unbalanceness and the coherence are affecting the Fisher information in the same mechanism. Therefore, the opposite variations of the two parameters are able to cancel each other's negative effects on the Fisher information, thus permitting a super-resolution. Fig~\ref{equivalent} (a), (b) illustrates specifically the equivalence of the two Fisher information behaviors. When the parameters are chosen appropriately, the two effects behave exactly the same as shown by same color lines in the two plots. In addition, the fact that angle $\alpha$ is controllable by the analyzer, one can always achieve finite FI for arbitrary unknown parameter $r$ as shown in Fig~\ref{equivalent}. This justifies that in the calculation of the Fisher Information (\ref{Ftot}), it is not necessary to estimate the unknown parameter $r$. 

Also, we observe that in the limit $s \rightarrow 0$, Fisher Information $F_{tot}(s,r)$ will never vanish for any unbalanced (i.e., $r\ne 1$) sources, see detailed proof in Supplement material section 4. From the equivalence of coherence and unbalanceness effect, it can therefore be concluded that by adjusting the coherence rotation angle $\alpha$ one can always avoid the balanced situation and thus avoid the vanishing Fisher Information at zero separation.

%The FI in the limit $s\rightarrow 0$ vanishes only in the case of the balanced amplitudes,  $\varphi=0 ,k\pi$ and $\alpha=\frac{k'\pi}{4}$, where $k=1,2,3...$ and $k'=1,3,5...$ (see the proof in Supplement 4).

{\em The third major message} of the Fisher information (\ref{Ftot}) is its counter intuitive decreasing behavior within the small separation regime, see illustrations in Fig.~\ref{FI} (a) and Fig.~\ref{equivalent}. Normally one would expect that as the separation $s$ of the two sources increases the Fisher information would also increase because it is natural to assume that larger separation means less relative error in measurements. However, as shown in Fig.~\ref{FI} (a) and Fig.~\ref{equivalent}, the Fisher information experiences a decrease and then an increase as the separation $s$ increases from zero. This behavior is due to the competing nature of coherence ($\alpha$) and unbalanceness ($r$) in affecting the Fisher information for any fixed relative phase $\varphi$ . 

This interesting behavior suggests the existence of a least resolvable separation $s_{\rm least}$ that leads to a minimum Fisher information $F_{tot}^{\rm min}$. All practical situations should avoid analyzing distances at the vicinity of this critical separation $s_{\rm least}$. To achieve this separation quantitatively, we analyze the derivative of the Fisher information (\ref{FI}). The vanishing derivative of $F_{tot}$ leads to the solutions of the following equation 
\begin{align}
\label{ali}
 \Lambda(s)e^{s^{2}/4\sigma^{2}}+\Pi(s)e^{s^{2}/8\sigma^{2}}+\Omega=0,
\end{align}
where, $\Lambda(s)=(a^{2}\cos^{2}\alpha+b^{2}\sin^{2}\alpha)(b^{2}\cos^{2}\alpha+a^{2}\sin^{2}\alpha)(1-\frac{2s^{2}}{8\sigma^{2}}), 
 \Pi(s)=[2 (a^{2}\cos^{2}\alpha+b^{2}\sin^{2}\alpha)-1](1-\frac{s^{2}}{8\sigma^{2}})ab\sin2\alpha\cos\varphi$ and
 $\Omega=-(ab\sin2\alpha\cos\varphi)^{2}$.
There are two trivial solutions $s=0$ and $s\rightarrow\infty$ as it is shown in Fig.~\ref{FI}(a) and  Fig.~\ref{equivalent}. The nontrivial solution can in general be always achieved numerically. For the commonly studied balanced case when $r=1$, the nontrivial solution of (\ref{ali}) can be obtained analytically as (see detailed analysis in supplemental material section 5)
\begin{align}
\label{minimum}
s_{\rm least}=\sigma\sqrt{4 + 4\mathcal{W}[-\frac{\sin^{2}2\alpha\cos^{2}\varphi}{e}]}
\end{align}
where $\mathcal{W}[.]$ is a special function known as the Lambert $\mathcal{W}$-function, which is an increasing function with a minimum at  $\mathcal{W}[\frac{-1}{e}]=-1$. When $\sin^{2}2\alpha\cos^{2}\varphi=1$, $s_{\rm least}=0$ which is exactly the case analyzed in \cite{Hradil2019O} for $\varphi=0,\pi$ and $\alpha=\pi /4$. 

Due to the equivalence of the coherence effect and unbalanceness effect as analyzed earlier, the least resolvable distance with respect to different values of the parameter $r$ can be obtained in exactly the same way as (\ref{minimum}), see also illustrations of $s_{\rm least}$ for different curves in Fig.~\ref{equivalent} (a) and (b). Since the detection basis (in terms of $\alpha$) can be controlled by the observer, all unbalanced cases can be treated equivalently as balanced cases but with a corresponding coherence angle $\alpha$. 

The least resolvable distance analysis provides an important guidance to avoid resolution of two-source separation at the vicinity of $s_{\rm least}$ in various practical situations.

\section{Conclusion}

To summarize, we have investigated sub-diffraction-limit resolution of two point-sources under two practical situations: arbitrary two-source unbalanceness and partial coherence. By including an entangled partner of the spatial property of the two sources to account the partial coherence, it is found that super-resolution can be achieved with high measurement estimation credibility (quantified by maximum Fisher information) even when the two-source separation reduces to zero. It is revealed that such achievement is due to the fact that the effect on Fisher information from partial coherence is equivalent to that of the two-source unbalanceness. Appropriate control of the rotated basis (i.e., adjustment of partial coherence) by the analyzer is able to counter effect arbitrary unbalanceness. Such a capability indicates that the realization super-resolution is independent of whether the unbalanceness and partial coherence are known or not. This justifies the exclusion of unknown parameters unbalancenss and partial coherence in analyzing Fisher information. 

We have also carried out a detailed analysis of the counter intuitive decreasing behavior of Fisher information as the two-source separation increases. This allows the discovery a characteristic equation to determine the ``least resolvable" distance. Analytical solutions in terms of Lambert W function are also achieved. Our results provide an important guidance for practical optical designs and engineering in the realization of optimum fine resolution.

{\bf Funding:} U.S. Army under contact No. W15QKN-18-D-0040.

{\bf Disclosures:} The authors declare no conflicts of interest

{\bf Acknowledgments:} We acknowledge partial financial support by Stevens Institute of Technology.

%%%%%%%%%%%%%%%%%%%%%%% References %%%%%%%%%%%%%%%%%%%%%%%%%

\end{document}